\documentclass[usenatbib]{mn2e}
\usepackage{graphicx}

\begin{document}
\topmargin-1cm

\newcommand\approxgt{\mbox{$^{>}\hspace{-0.24cm}_{\sim}$}}
\newcommand\approxlt{\mbox{$^{<}\hspace{-0.24cm}_{\sim}$}}
\newcommand{\be}{\begin{equation}}
\newcommand{\ee}{\end{equation}}
\newcommand{\bea}{\begin{eqnarray}}
\newcommand{\eea}{\end{eqnarray}}
\newcommand{\lexp}{\mathop{\langle}}
\newcommand{\rexp}{\mathop{\rangle}}
\newcommand{\rexpc}{\mathop{\rangle_c}}
\newcommand{\nbar}{\bar{n}}
\newcommand{\zmax}{Z_{\rm max}}
\newcommand{\amass}{10^{13} \,M_\odot}
\newcommand{\hmass}{10^{14} \,M_\odot}
\newcommand{\mmin}{M_{\rm{min}}}

\title{Dark matter halo abundances, clustering and assembly histories at
high redshift}
\author[Cohn \& White]{J.D. Cohn${}^{1}$ and Martin White${}^{2}$\\
${}^1$Space Sciences Laboratory,\\ 
${}^2$Departments of Physics and Astronomy,\\
University of California, Berkeley, CA 94720}

\date{\today}
\maketitle

\begin{abstract}
We use a suite of high-resolution N-body simulations to study the properties,
abundance and clustering of high mass halos at high redshift, including their
mass assembly histories and mergers.
We find that the analytic form which best fits the abundance of halos depends
sensitively on the assumed definition of halo mass, with common definitions of
halo mass differing by a factor of two for these low concentration, massive
halos.  A significant number of massive halos are undergoing rapid mass
accretion, with major merger activity being common.  We compare the mergers
and mass accretion histories to the extended Press-Schechter formalism.

We consider how major merger induced star formation or black hole accretion
may change the distribution of photon production from collapsed halos, and
hence reionization, using some simplified examples.  In all of these, the
photon distribution for a halo of a given mass acquires a large scatter.
If rare, high mass halos contribute significantly to the photon production
rates, the scatter in photon production rate can translate into additional
scatter in the sizes of ionized bubbles.
\end{abstract}

\section{Introduction}

Observations of the anisotropy of the cosmic microwave background (CMB)
radiation have given us unprecedented knowledge of the very early Universe and
dramatically confirmed the picture of large-scale structure as arising from
the gravitational amplification of small perturbations in a Universe
with a significant cold dark matter component \citep{Smo92}.
In this model the ionization history of the Universe has two main events,
a `recombination' at $z\sim 10^3$ in which it went from ionized to neutral
and a `reionization' during $z\sim 7-12$ in which the radiation from early
generations of collapsed objects was able to ionize the intergalactic medium.
The former event is strongly constrained by the CMB.
A new generation of instruments will soon allow us to probe this second
event: ``the end of the dark ages''
\citep[for reviews of reionization see e.g.][]{BarLoe01,CooBar05,
FanCarKea06,FurOhBri06}.

Since at reionization a very small fraction of the mass affected each and
every baryon in the Universe, reionization is particularly sensitive to
the distribution and behavior of collapsed structure.
We expect that the ionizing sources are situated in large
($T_{\rm vir}>10^4$K or $M>10^7\,h^{-1}M_\odot$)
dark matter halos where the gas can cool efficiently to form stars\footnote{We
will only consider Pop II stars here; Pop III stars, which can form in the
absence of metals in smaller halos, are expected to be less likely by 
redshift 10 \citep{Yos04}.}.
Models for the sources of reionization thus often start with estimates of the
number and properties of virialized dark matter halos at high redshift,
the focus of this paper. At $z=10$, halos with
$M>10^{9}\,h^{-1}M_\odot$
are expected to be biased similarly to very massive clusters
($M>10^{15}\,h^{-1}M_\odot$) today, with the most massive and recently
formed halos growing rapidly and merging frequently.
We explore some properties of these collapsed halos at a high redshift using
a suite of high resolution, collisionless, N-body simulations.
We pay particular attention to merger rates and mass accretion histories with
an eye to applications for reionization.
We also compare the N-body results with the predictions of the oft-used
\citet{PreSch74} formalism.

\begin{figure*}
\begin{center}
\resizebox{3.2in}{!}{\includegraphics{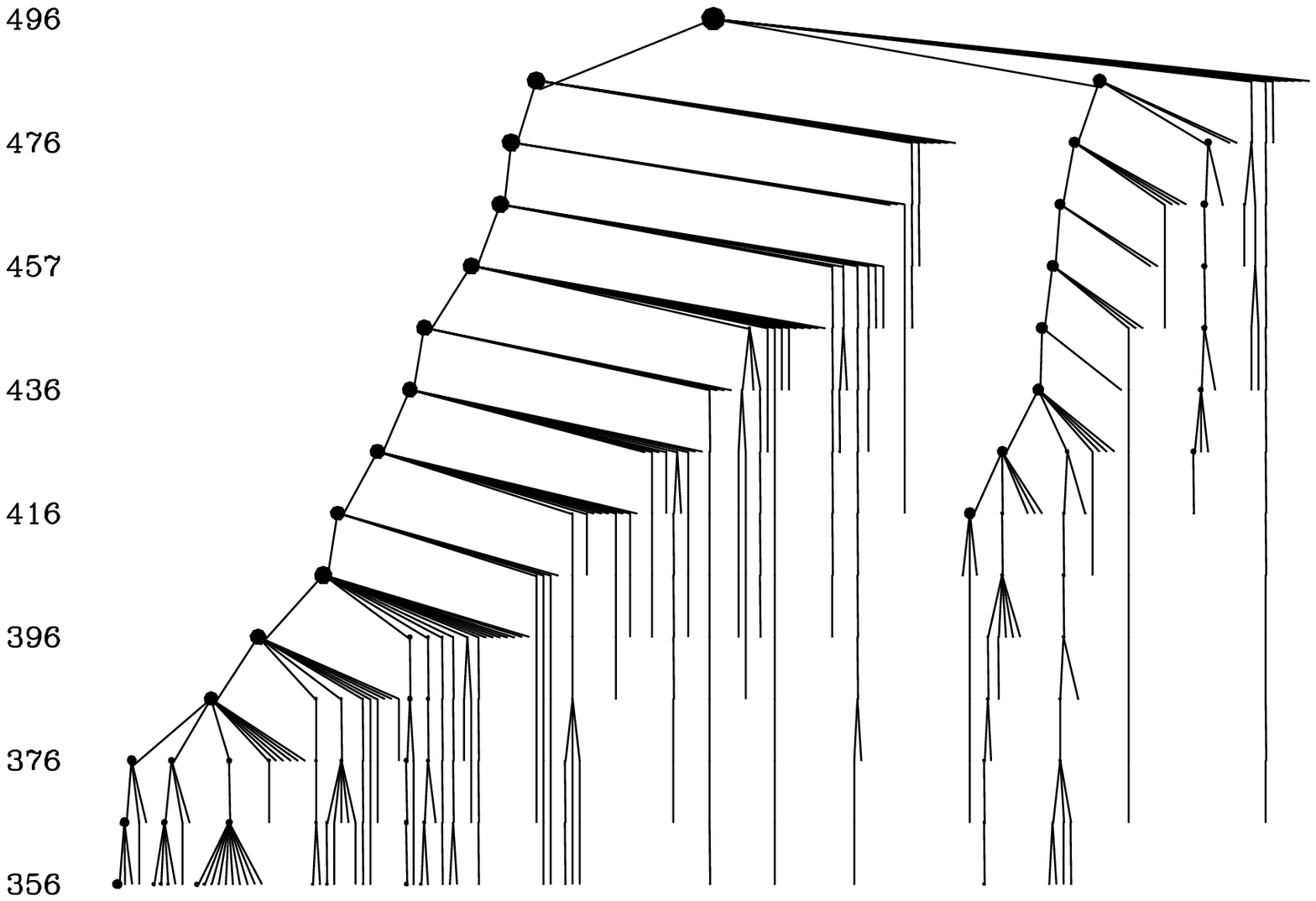}}
\resizebox{3.2in}{!}{\includegraphics{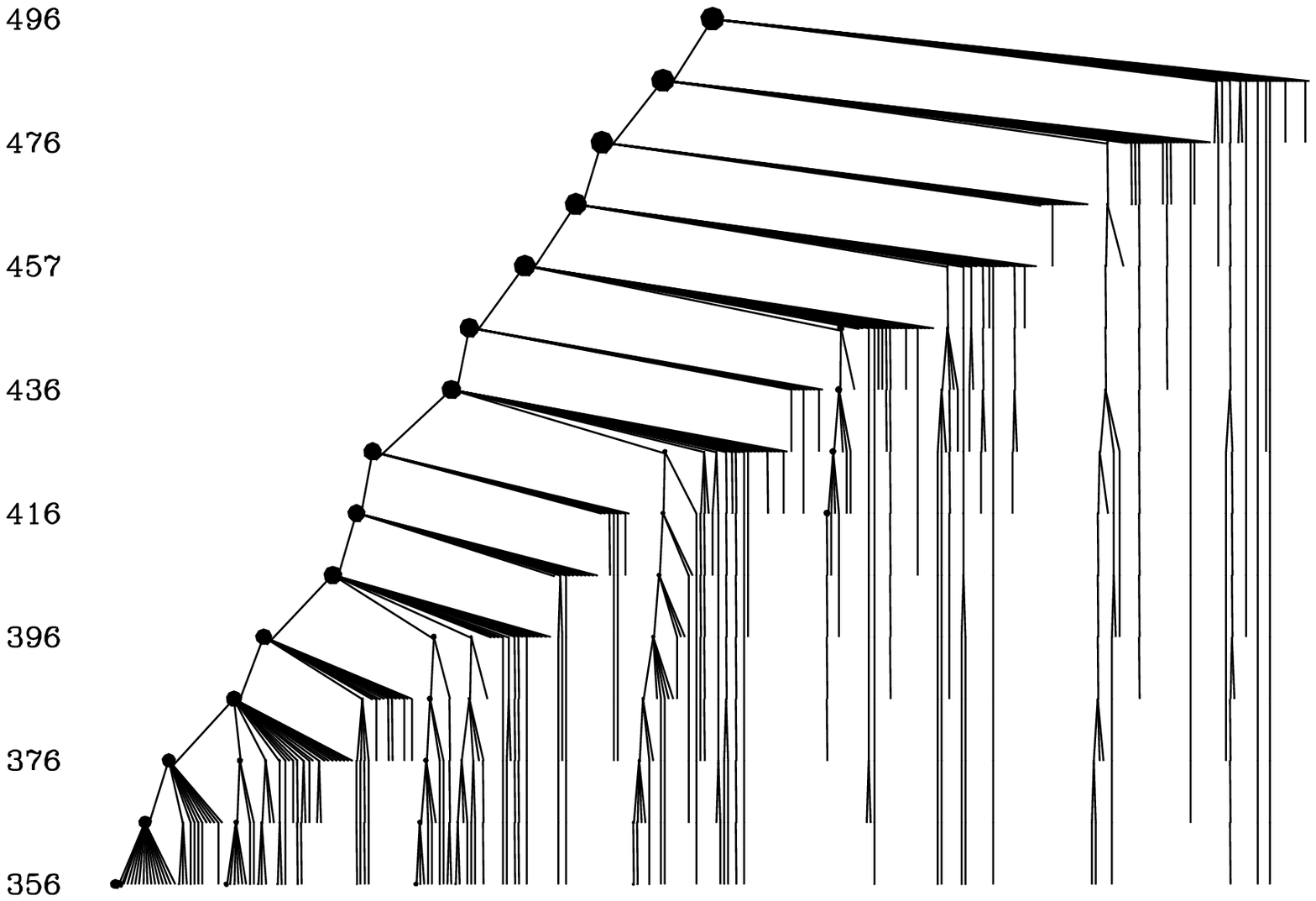}}
\end{center}
\caption{Illustrative merger trees for two halos with masses of $1.2$ (left)
and $2.9\times 10^{10}\,h^{-1}M_\odot$ (right).  
Time runs upwards in steps of 10 Myr, from $z=12.7$ (bottom) to $z=10$ (top)
and the age of the Universe (in Myr) is shown at every second step.  At each
time the area of the symbol is proportional to the halo mass, with masses
decreasing to the right in each group, and lines show the progenitor
relationship.  The leftmost branch shows the main trunk of the tree.
The halo at left has a (major) 1:2 merger at the last time step, while the
main trunk of halo at right has a 1:2 merger at the first time step, a 1:6
two steps later and then only smaller mergers after that.}
\label{fig:treeplot}
\end{figure*}

If halo mergers are accompanied by a temporary increase in photon production
(due either to starbursts or increased black hole accretion
e.g.~\citealt{Car90,BarHer91,BarHer96,MihHer94,MihHer96,KauHae00,CavVit00})
we expect reionization to be influenced by the merger and accretion history
of dark matter halos, beyond just the fact that more massive halos
emit more photons.  With a simple model of star formation we show that
merger-induced scatter in photon production may be significant, with the
production rates acquiring a substantial tail to large photon production rates.
Since the massive halos are relatively rare, this individual halo scatter
is expected to translate into a scatter of photon production rates inside 
ionized regions, changing the bubble distribution.

The outline of the paper is as follows.  In \S\ref{sec:sim} we describe
the N-body simulations.  The basic halo properties are described in
\S\ref{sec:halo} along with the results for mergers and mass gains and
the comparison to Press-Schechter.
The consequences of this merging in a simple model for photon production
are elucidated in \S\ref{sec:reion} and we summarize and
conclude in \S\ref{sec:conclusions}.

\section{Simulations and parameters} \label{sec:sim}

We base our conclusions on 5 dark matter only N-body simulations
of a $\Lambda$CDM cosmology with $\Omega_m=0.25$, $\Omega_\Lambda=0.75$,
$h=0.72$, $n=0.97$ and $\sigma_8=0.8$, in agreement with a wide array of
observations.
The initial conditions were generated at $z=300$ using the Zel'dovich
approximation applied to a regular, Cartesian grid of particles.
Our two highest resolution simulations employed $800^3$ equal mass
particles ($M=2\times 10^6$ and $1.7\times 10^7\,h^{-1}M_\odot$) in
boxes of side $25$ and $50\,h^{-1}$Mpc with Plummer equivalent smoothings
of $1.1$ and $2.2\,h^{-1}$kpc.  They were evolved to $z=10$ using the
{\sl TreePM\/} code described in \citet{TreePM}
\citep[for a comparison with other codes see][]{CodeCompare}.
We ran 3 additional, smaller simulations in a $20\,h^{-1}$Mpc box, one
with $600^3$ particles and two with $300^3$ particles (each started at
$z=200$).  A comparison of the boxes allows us to check for finite volume,
finite mass and finite force resolution effects.  We shall comment on
each where appropriate.

The phase space data for the particles were dumped at $15$ outputs spaced
by $10\,$Myr from $z=12.7$ to $z=10$ for all but the largest
box.  The lower resolution of the largest box makes it less
useful for merger trees, so it was sampled for only subset of 
these output times, ending at $z=10$.  For each output we generate a
catalog of halos using the Friends-of-Friends (FoF) algorithm \citep{FoF}
with a linking length, $b$, of $0.168$ times the mean inter-particle spacing.
This partitions the particles into equivalence classes, by linking together
all particle pairs separated by less than $b$.
The halos correspond roughly to particles with $\rho>3/(2\pi b^3)\simeq 100$
times the background density.
We also made catalogs using a linking length of $0.2$ times the mean
inter-particle spacing, which we shall discuss further below.
We found that the FoF algorithm with a larger linking length had a tendency
to link together halos which we would, by eye, have characterized as separate
\citep[see also][for similar discussion]{FoF,ColLac96}.
This problem is mitigated with our more conservative choice of $b$.

For each halo we compute a number of properties, including the potential
well depth, peak circular velocity, the position of the most bound particle
(which we take to define the halo center) and $M_{180}$, the mass interior
to a radius, $r_{180}$, within which the mean density is $180$ times the
background density\footnote{Note this is simply a definition of halo mass,
not the halo finder.  We still use FoF particles to define the group centers.
However given the center we use all of the particles in the simulation
when determining $M_{180}$.  Our $M_{180}$ masses should thus be comparable
to the sum of the particles in an $SO(180)$ group -- a common definition that 
employs both the $SO(180)$ halo finder and definition
of mass.}.
As discussed in \citet{Whi01,TreePM} and \citet{HuKra03}, the choice of halo
mass is problematic and ultimately one of convention.  We shall return to
this issue in the next section.

Merger trees are computed from the set of halo catalogs by identifying for
each halo a ``child'' at a later time.  The child is defined as that halo
which contains, at the later time step, more than half of the particles in
the parent halo at the earlier time step (weighting each particle equally).
For the purposes of tracking halos this simple linkage between outputs
suffices (note that we do not attempt to track subhalos within larger halos,
which generally requires greater sophistication).  Two examples of the halo
merger trees are given in Fig.~\ref{fig:treeplot}, where we see a rich set
of behaviours, including major and minor mergers and many body mergers.  
From the merger trees it is straightforward to compute the
time when a halo `falls in' to a larger halo, the number and masses of the
progenitors etc. 

Due to finite computational resources, all N-body simulations must trade-off
computational volume for mass resolution.  By running multiple simulations we
can overcome this to some extent, but not entirely.  We have chosen to slightly
under-resolve the low mass ($T_{\rm vir}\simeq 10^4$K) halos in order to
simulate a slightly larger volume, since our focus will be on the more
massive halos which have more frequent major mergers.
Under reasonable assumptions (see below) between $\frac{1}{3}-\frac{2}{3}$
off all photon production occurs in halos more massive than
$10^9\,h^{-1}M_\odot$ at $z=10$, and we easily resolve these objects with
the $25\,h^{-1}$Mpc simulation which we use for the bulk of the paper.

\section{Halo properties} \label{sec:halo}

\subsection{Halo abundance and clustering}

\begin{figure}
\begin{center}
\resizebox{3.1in}{!}{\includegraphics{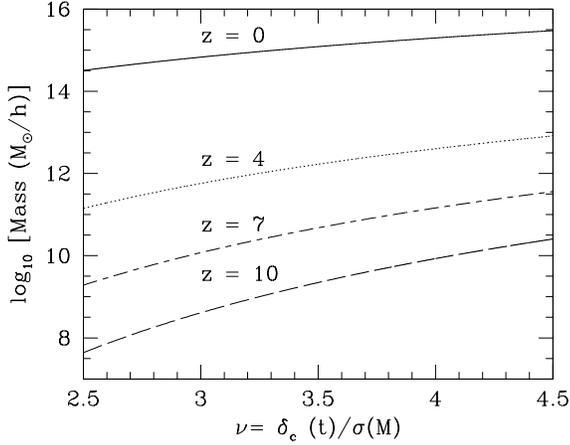}}
\end{center}
\caption{The peak height, $\nu$ -- which governs the abundance, clustering
and merging behavior in analytic models -- for $z=0$, 4, 7 and 10.
For example, objects with $\nu=3$ have $M\simeq 4\times 10^8\,h^{-1}M_\odot$
at $z=10$ but $M\simeq 6\times 10^{14}\,h^{-1}M_\odot$ at $z=0$.}
\label{fig:comparenu}
\end{figure}

The highest mass objects in our volume have mass $\sim 10^{10} h^{-1}M_\odot$
and radii of several tens of kpc.
At $z=10$ these halos are analogous to rich clusters today, being recently
formed and rare: Fig.~\ref{fig:comparenu} shows the mass as a function of
peak height, $\nu\equiv\delta_c/\sigma(M)$, at $z=10$, 7, 4 and 0.
The threshold $\delta_c(t)$ is defined as $1.686/D(t)$, where $D$ is the
linear growth factor normalized to unity at $z=0$ and $\sigma^2(M)$ is
the variance of the mass computed using linear theory at $z=0$.  In our
cosmology $\delta_c(z=10)\simeq 13.8$.
Due to the flatness of the dimensionless power on the scales of interest,
the slightly red initial spectrum and the low clustering amplitude, the
characteristic mass, $M_\star$, where $\sigma\sim\delta_c$, is
${\mathcal O}(1)M_\odot$ at $z=10$, so all of the halos we consider are 
$\gg M_\star$.

\begin{figure}
\begin{center}
\resizebox{3.1in}{!}{\includegraphics{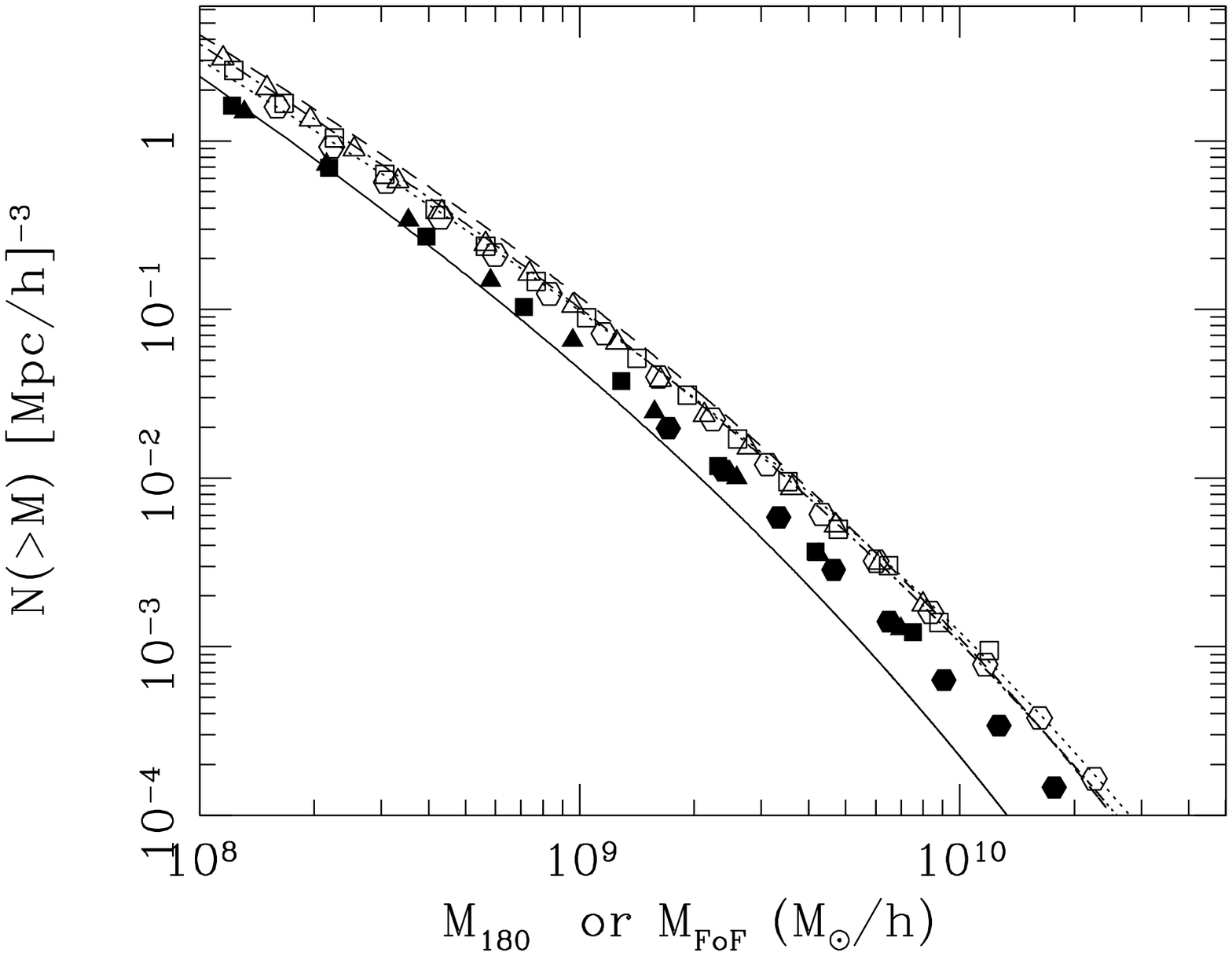}}
\end{center}
\caption{The mass functions for our box compared to the
\protect\citet{War05} (dashed line),  
\protect\citet{Jen01} (dash-dotted line), \protect\citet{SheTor99}
(dotted line) and Press-Schechter (solid lower line) mass functions
for our cosmology.  N-body $M_{180}$ results are plotted as solid symbols for
the $800^3$ ($25 h^{-1}$Mpc, squares, $50 h^{-1}$Mpc, hexagons) 
and $600^3$ ($20 h^{-1}$ Mpc, triangles) runs.  Only masses where there
are more than 10 halos in the box and where resolution effects are unimportant
are shown.  Open symbols denote the analogous
FoF(0.2) mass functions for the same simulations.}
\label{fig:nofm}
\end{figure}

One of the most basic and useful quantities we can derive from the simulations
is the mass function, the spatial abundance of halos as a function of mass.
High redshift mass functions have been studied by many groups
\citep[e.g.][]{JanHer01,Ree05,Spr05,Ree07,Hei06,TraCen06,Ili05,Ili06a,Mai06,
Zah07,Luk07} and \citet{Luk07} offer a comprehensive summary of recent work.
Most previous work finds mass functions which are better fit by the
\citet{SheTor99}, \citet{Jen01} or \citet{War05} form.  We find that the
appropriate mass function to use depends primarily on the definition of mass
chosen and definitions which at $z\simeq 0$ give very similar mass functions
can give quite different ones at $z=10$.

We show the mass function(s) from our three highest resolution simulations
in Fig.~\ref{fig:nofm}.  If we use as our mass estimator the sum of the
particle masses in the FoF(0.2) groups (open symbols) then we find good
agreement with the \citet{SheTor99} or \citet{Jen01} forms.
This is the procedure followed by most of the groups above\footnote{Due to
the finite size of our boxes the mass function is slightly suppressed at
high mass.  We can estimate this suppression using (extended) Press-Schechter
theory, assuming we have simulated the conditional mass function within a
region of exactly mean density on the mass scale of the box.
The mass functions plotted have been corrected for this expected suppression,
which ranges from $<1\%$ to 22\% over the mass range plotted.
See, e.g.~\protect\citet{Luk07,Ree07} for further discussion.}.
However if we choose instead to use $M_{180}$ as our mass estimator (filled
symbols) we find a different mass function.  
Although this mass function shows a marked excess of high mass halos compared
to the \citet{PreSch74} form, it is a better fit than the alternate forms
mentioned above.   
Agreeably, for the scales plotted, the $M_{180}$ mass function is independent
of the initial FoF group catalog used to define the centers about which
$M_{180}$ is determined.
This is not too surprising as the group centers hardly change and the number
of ``small'' groups which split off of larger FoF groups as the linking length
is decreased is tiny compared to the number of low-mass ``field'' halos.
The differences in mass functions then comes primarily from the definition of
the masses of the found objects.
Comparing halo by halo the FoF(0.2) masses are almost twice $M_{180}$, though
the difference depends on mass.
A similar difference was also noted by \citet{Ree07} as a shift to lower
abundance at fixed mass when comparing an FoF(0.2)-based mass function to
that of a different halo finder.  We believe the primary issue is not the halo
finder, but the mass definition.  Their second halo finder assigns masses
which are essentially our $M_{180}$.
The mass discrepancy is much larger for these halos at $z=10$ than it is for
group and cluster-sized halos at the present day
\citep[e.g. Figure 11 in][]{TreePM}.

\begin{figure}
\begin{center}
\resizebox{3.1in}{!}{\includegraphics{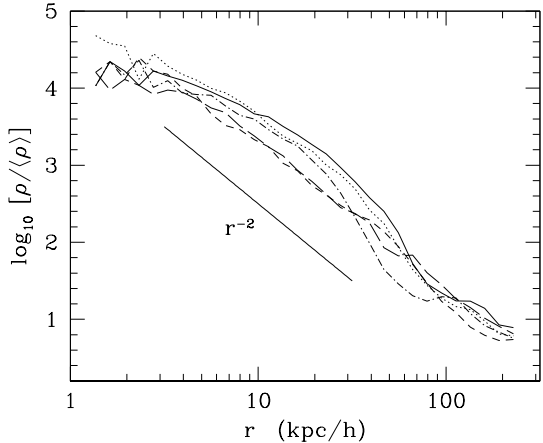}}
\end{center}
\caption{Density profiles of the 5 most massive halos in the $800^3$ run
at $z=10$.  The masses range from $1-3\times 10^{10}\,h^{-1}M_\odot$.
The 2 halos with the flatter profiles (short and long dashed lines)
correspond to the $3^{\rm rd}$ and $4^{\rm th}$ most massive halos and both
have had a major merger (greater than 1:6) within the previous $10\,$Myr.
Halo sizes ($r_{180}$) are below $100\,h^{-1}$kpc for all of the halos shown.
The solid line, offset, shows an isothermal sphere profile
($\rho\propto r^{-2}$) for comparison.  This indicates why 
FoF(0.2) masses assuming an isothermal profile may be
expected to disagree with SO(180) masses as discussed in the text.
}
\label{fig:profile}
\end{figure}

The mass differences are quite interesting.  The historical argument 
for choosing
FoF(0.2) was that the FoF group finder selects particles approximately within
a density $3/(2\pi b^3)\simeq 60$ times the mean density.
For a singular isothermal sphere profile ($\rho\propto r^{-2}$) and a critical
density Universe the mean enclosed density is thus $180\,\rho_{\rm crit}$,
in accord with arguments based on spherical top-hat collapse
\citep[e.g.][]{Pea98}.
At $z=10$ the Universe is close to critical density so we might expect the
FoF(0.2) and $M_{180}$ mass functions to agree better than at lower $z$ where
$180\bar{\rho}\simeq 45\rho_{\rm crit}$.  However, we are focusing on very
high mass halos which have only recently formed at $z=10$.
They are therefore less centrally concentrated\footnote{They correspond roughly
to $c\simeq 2-5$ for halos of the form proposed by \protect\citet{NFW}.}
than a `typical' halo.    This can also be seen in Fig.~\ref{fig:profile},
where halo profiles are compared to the isothermal sphere profile.  As
the halo profiles are less steep than the isothermal sphere form assumed in the
argument above, this leads to the differences in
mass between the FoF(0.2) and $SO(180)$ definitions.

By contrast, we find that the FoF(0.168) mass function is very similar to the
$M_{180}$ points plotted, and a halo by halo comparison shows that the two
masses agree to within 20-30 per cent.  
As we go down the mass function, to more concentrated, lower mass halos, we
expect FoF(0.2) to better match $M_{180}$ \citep[e.g.][]{ColLac96}.

In general, given the strong dependence of the mass function upon the mass
definition, and the ambiguity in this quantity in many analytic treatments,
significant care must be taken when making predictions for the abundance of
halos.
Even if we decide to treat all halos as a simple 1-parameter family, it is
likely preferable to make comparisons with some quantity more directly related
to observables (such as circular velocity, halo virial temperature or potential
well depth) or to discuss statistics as a function of number density rather
than mass.

\begin{figure}
\begin{center}
\resizebox{3.1in}{!}{\includegraphics{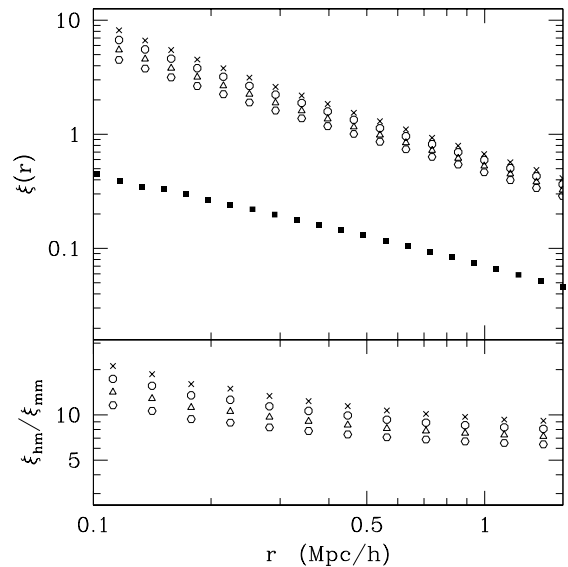}}
\end{center}
\caption{(Top) The halo-dark-matter cross correlation, $\xi_{hm}(r)$ (top),
for halos with (comoving) number density $10^{-2.0}$, $10^{-1.5}$,
$10^{-1.0}$ and $10^{-0.5}\,h^3{\rm Mpc}^{-3}$ (open symbols from top to
bottom) from our $50\,h^{-1}$Mpc simulation.  The solid squares show the dark
matter correlation function, $\xi_{mm}(r)$.  The ratio,
$b(r)\equiv\xi_{hm}(r)/\xi_{mm}(r)$, is shown in the lower panel.}
\label{fig:bias}
\end{figure}

Like rich clusters we expect that these massive halos, in the process
of formation, will not lie on the usual `vacuum' virial relation 2KE$=$PE,
where KE and PE refer to the potential and kinetic energy respectively.
In fact we find that 2KE/PE$\simeq 1.4$ for halos in the range
$10^8-10^{10}\,h^{-1}M_\odot$, very similar to the value found for rich
clusters today \citep{KneMul99,CohWhi05,Sha06}.
A similar `excess' kinetic energy was also found by \citet{JanHer01} for
lower mass halos.  The ratio is larger than unity because of the steady
accretion of material onto the cluster \citep{ColLac96}.

Fig.~\ref{fig:bias} shows the clustering of the dark matter and the halos
{}from our $50\,h^{-1}$Mpc run.
We plot the auto-correlation function of the dark matter and the
cross-correlation of the halo centers with the dark matter respectively.
The latter is both less subject to noise\footnote{There is essentially no
shot-noise for the dark matter, $\xi_{mm}(r)$, and jackknife errors on the
cross-correlation, $\xi_{hm}$, are a few percent for the samples shown.
Jackknife drastically underestimates the errors from finite volume however.}
{}from our small sample of massive halos and more applicable to understanding
how radiation from the halos would influence the surrounding mass.
The ratio of the cross- to auto-correlation functions defines the scale
dependent bias, $b_h(r)$.

\begin{figure}
\begin{center}
\resizebox{3.1in}{!}{\includegraphics{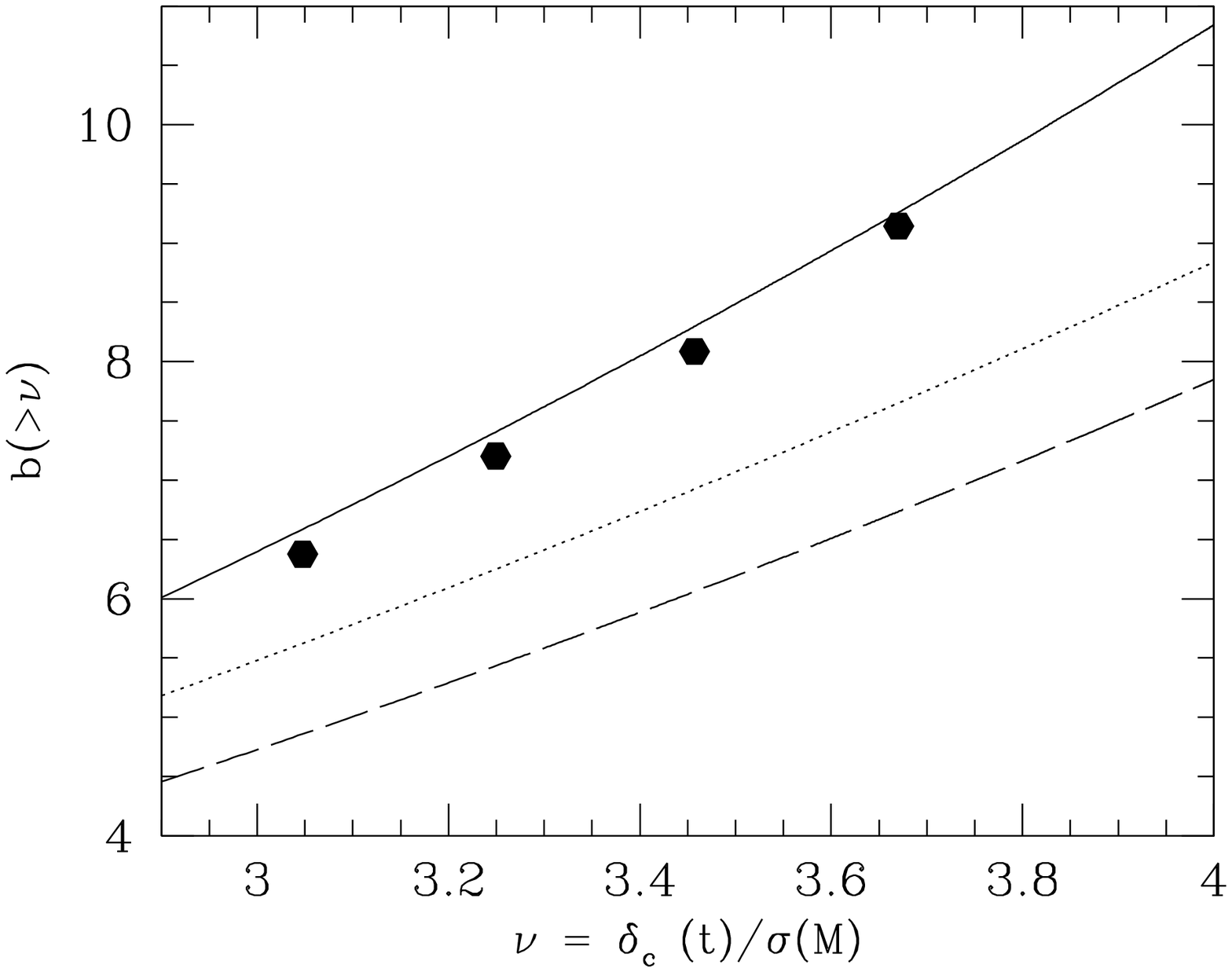}}
\end{center}
\caption{A comparison of the large-scale bias measured for the mass
thresholded samples of Fig.~\protect\ref{fig:bias} with a number of
theoretical models: the bias of the Press-Schechter mass function
\protect\citep[as computed by][solid]{Efs88,ColKai89},
the Sheth-Tormen mass function \protect\citep[][dashed]{SheTor99} and
the fitting function of \protect\citet[][dotted]{SheMoTor01}.
Although the mass function is in good agreement with that of
\protect\citet{SheTor99}, their bias formula underestimates the clustering
of the rarest halos.}
\label{fig:b_vs_nu}
\end{figure}

The mass auto-correlation function is in good agreement between the $25$
and $50\,h^{-1}$Mpc boxes up to $1\,h^{-1}$Mpc, with $\xi$ from the
$25\,h^{-1}$Mpc box falling below that of the $50\,h^{-1}$Mpc box beyond
this scale.
The $20\,h^{-1}$Mpc box has noticeably less power over a wide range
of scales.
For the masses where we can compare and for the range of linear scales plotted,
the halo-mass cross-correlation functions of the $25$ and $50\,h^{-1}$Mpc boxes
are in excellent agreement, so we have shown the results only for the
$50\,h^{-1}$Mpc box.

Our halo samples are mass thresholded, however by using number density
as our marker we largely avoid the issues of mass definition discussed
earlier.  The differences in bias at fixed $\bar{n}$ for the different
mass choices, arising from the scatter between different mass definitions,
is only a few percent.
Taking $b_h(1.5\,h^{-1}{\rm Mpc})$ as the asymptotic value, the large-scale
bias is in good agreement with the models of
\citet{PreSch74,Efs88,ColKai89,MoWhi96,Jin98}
and $\sim 30\%$ higher than that of \citet{SheTor99}.
Those of \citet{SheMoTor01} and \citet{Tin05} lie in between.
(The model of \citet{SelWar04} only extends up to 100 times the non-linear
mass, where $b\sim 3$, and it not applicable to our results.)
To make contact with the earlier literature we plot in Fig.~\ref{fig:b_vs_nu}
the bias as a function of peak height, $\nu$, obtained from $\bar{n}$ using
the \citet{SheTor99} mass function.
When computing $b(>\nu)$ in the simulation we rank order the halos by FoF(0.2)
mass in order to best match the chosen mass function.  This mass function is
then used when analytically computing the halo-weighted bias $b(>\nu)$ from
each of the analytic forms which provide $b(\nu)$.  Because of this the
\citet{SheTor99} bias function is the only one which would give an average
bias of unity when integrated over $\nu$.

Similar trends for rare halos to have larger bias than the modern fits predict
have been seen at lower redshift
\citep[e.g.][for recent work]{Shen08,WhiMarCoh08,AngBauLac08}
but we must also remember that $b_h(1.5\,h^{-1}{\rm Mpc})$ is likely higher
than $b_h(r\to\infty)$ so the degree of overshoot is hard to quantify
precisely.
As expected, the clustering strength is an increasing function of mass
\citep{Kai84,Efs88,ColKai89}, or a decreasing function of halo abundance.

\subsection{Mergers and Mass Gains} \label{sec:merge}

\begin{figure}
\begin{center}
\resizebox{3.1in}{!}{\includegraphics{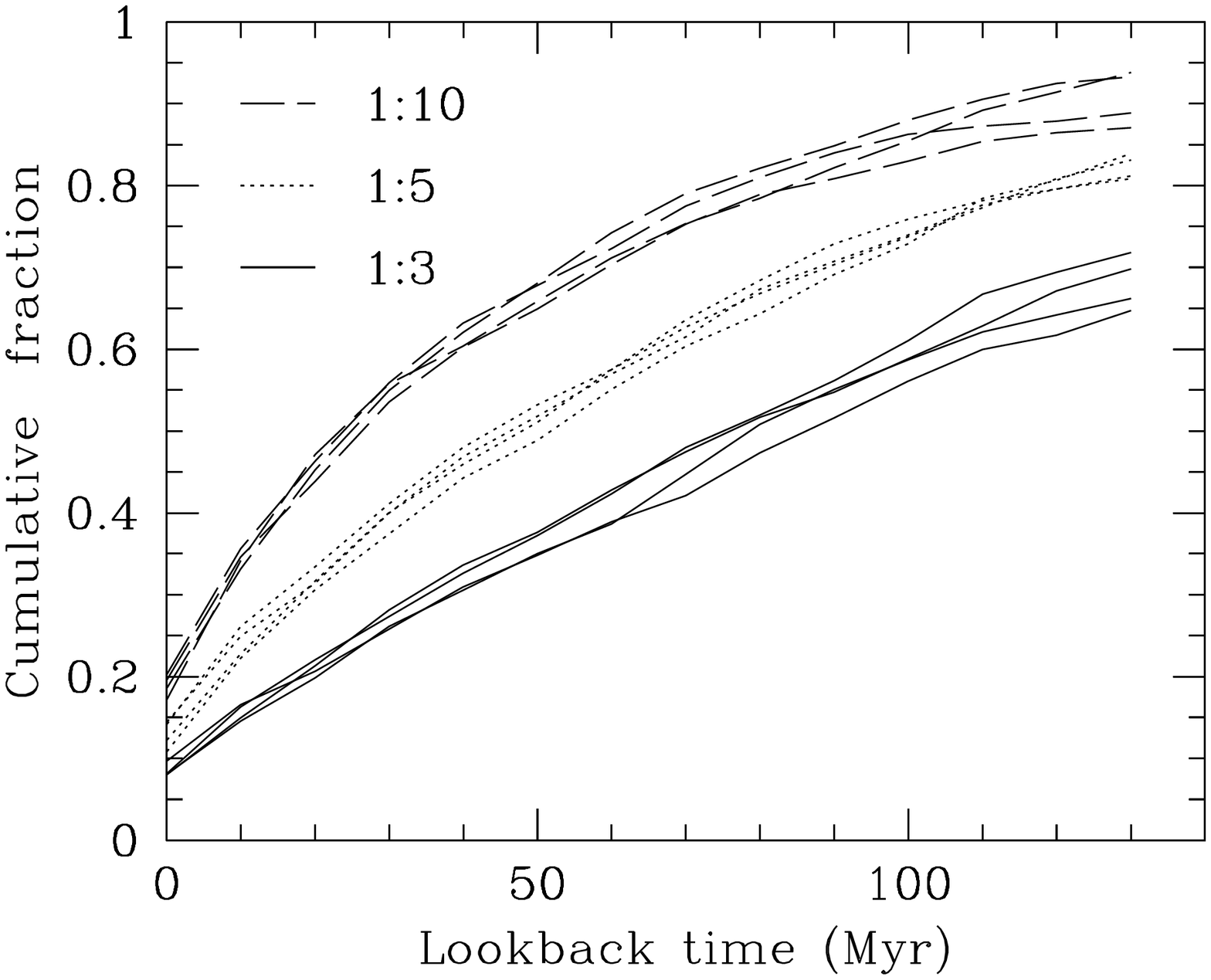}}
\resizebox{3.1in}{!}{\includegraphics{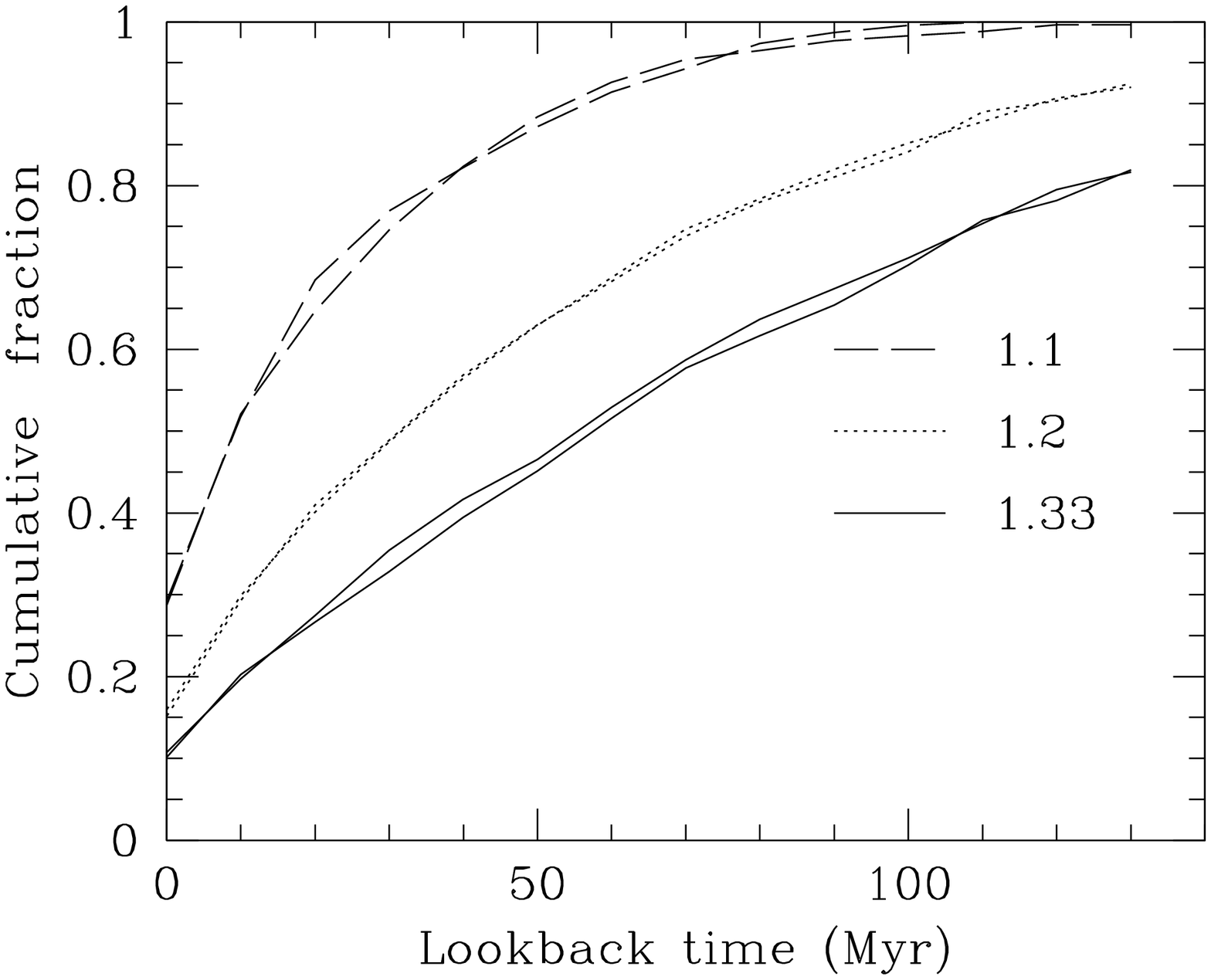}}
\end{center}
\caption{(Top) The fraction of halos with $M>10^9\,h^{-1}M_\odot$ which
have had a 1:10, 1:5 or 1:3 merger (top to bottom) back to the lookback
time shown in all 4 of our simulations.
(Bottom) The fraction of halos with $M>10^9\,h^{-1}M_\odot$ which have a
large mass gain ($m_f/m_i\geq 1.1$, 1.2, 1.33) vs.~time.  Here we show
only the two highest resolution simulations for clarity.
Both plots would coincide if all mergers were 2-body within $10\,$Myr.}
\label{fig:mrates}
\end{figure}

We now consider the hierarchical assembly of the dark matter halos through
merging and accretion.  We shall use the $800^3$, $25\,h^{-1}$Mpc simulation
since it provides both high mass resolution and a representative volume.
Since our progenitor relationships are based on particles in the FoF groups,
we use the FoF(0.168) masses for consistency.
As discussed earlier, for our massive halos these masses are within $20-30$
per cent of $M_{180}$ and none of our conclusions depend sensitively on this
choice.  Fig.~\ref{fig:mrates} shows the fraction of halos with
$10^9\le M\le 10^{10}\,h^{-1}M_\odot$ which have experienced at least
one major merger as function of lookback time, in intervals of $10\,$Myrs.
We show three different definitions of `major' merger, where the largest
two progenitors of the halo have ratios below 1:10, 1:5 or 1:3.
Mergers are frequent but not ubiquitous -- not all halos have had a major
merger within $140\,$Myrs, but many have.  The fraction decreases for
smaller mass ratios and for lower mass halos, as expected.

We can also consider mass gains between time steps, often denoted in the
literature as $m_f/m_i$ where $m_i$ is the mass of the largest progenitor
at the earlier time and $m_f$ is the mass of the halo under consideration.
Mass gains are sometimes used as a proxy for mergers.
Fig.~\ref{fig:mrates} shows those halos whose mass increased by at least
10, 20 or 33 per cent as a function of lookback time.  The top and bottom
panels of Fig.~\ref{fig:mrates} would be identical if all mergers were two
body and there was no smooth accretion.  As can be seen in 
Fig.~\ref{fig:treeplot} this is not the case; 
Fig.~\ref{fig:mrates} quantifies this difference for major mergers.

\begin{figure}
\begin{center}
\resizebox{3.1in}{!}{\includegraphics{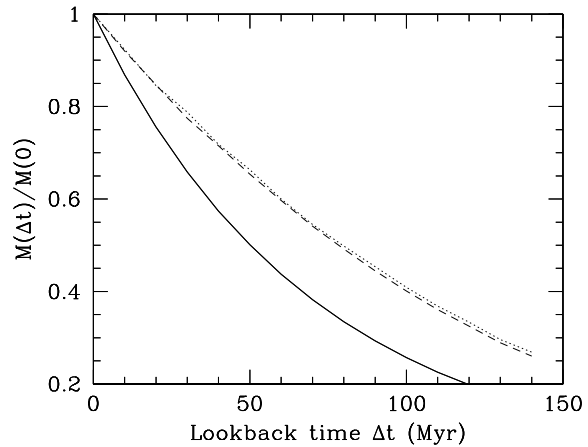}}
\end{center}
\caption{The mass accretion history for halos in the range
$(5-8)\times 10^8\,h^{-1}M_\odot$ from the $800^3$ (dashed) and $600^3$
(dotted) simulations and the functional form of \protect\citet[][solid]{Mil06}
based on EPS.}
\label{fig:mah}
\end{figure}

The Press-Schechter model predicts the evolution of the mass function, and
it can be extended to make predictions for the time history of halos.
This ``excursion set formalism'' is often called extended Press-Schechter
\citep{bondetal,Bow91,LacCol93,LacCol94,KitSut96}
and denoted EPS -- see \citet{Zen06} for a recent review.
Although it is analytically tractable, it has many inconsistencies and does
not compare particularly well to N-body simulations
\citep[see e.g.][]{ShePit97,Tor98,Som00,CohBagWhi01,BenKamHas05,Li07}.
For example, \citet{Li07} found that with EPS halos of mass
$10^{11}-10^{14}\,h^{-1}M_\odot$ at $z\simeq 0$ formed later than in
N-body simulations \citep[but see][for a slightly different quantity]{PMP00}.
In Fig.~\ref{fig:mah} we compare the N-body mass accretion histories for
massive halos at $z=10$ to a model by \citet{Mil06} based on EPS which
predicts almost exponential growth with redshift.  (Other analytic models
also exist, see e.g.~\citet{NeivanDek06} for a summary and comparison, 
there are some discrepancies between these which are not yet fully understood.)
We find that EPS predicts mass growth which is too rapid also
for the high mass, high redshift regime studied here.
The N-body mass accretion histories {\it are\/} relatively well fit by an
exponential in $z$ -- a growth model also proposed by \citet{Wec02} on the
basis of N-body simulations of galaxy-sized halos at low $z$ -- but the
coefficient predicted by \citet{Mil06} is larger than measured in the
simulations.

Perhaps the most common use of EPS is to predict merger rates, and EPS has
been used in this context in several recent models of reionization.
To compare the EPS predictions with our simulations we computed merger rates
using only our last ($10\,$Myr) time step, taking for any halo with $z=10$ mass
within $M_f$ to $M_f+\Delta M_f$ the distribution of progenitors\footnote{We
thank D.~Holz for suggesting this as a useful comparison quantity.},
$M_{\rm prog}$.  The EPS prediction can be found in the Appendix.

\begin{figure}
\begin{center}
\resizebox{3.1in}{!}{\includegraphics{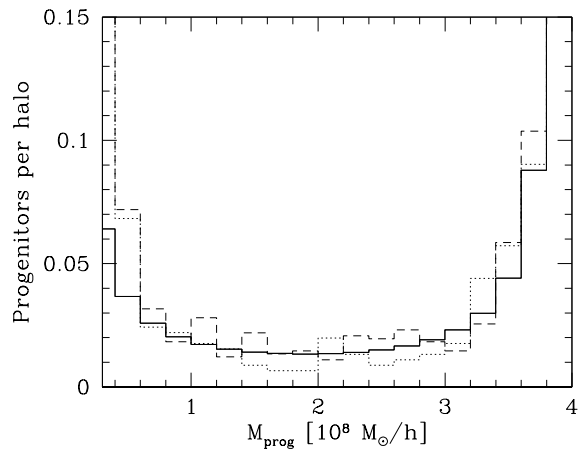}}
\end{center}
\caption{EPS (solid) and simulation (dotted and dashed) results for the
number of progenitors of halos with $M_f=(4-4.5)\times 10^8\,h^{-1}M_\odot$
as a function of $M_{\rm prog}$.}
\label{fig:epscomp}
\end{figure}

We show a representative example of $N(M_{\rm prog})/N(M_f)$ for $M_f$ in
the range $(4-4.5)\times 10^8\,h^{-1}M_\odot$ in Fig.~\ref{fig:epscomp}.
For most of the range the agreement is reasonably good.  At the low mass
end EPS significantly underpredicts the number of predecessors found in
our simulations \citep[see also][]{Per01}.
At the high mass end the EPS rate starts to climb rapidly, eventually
diverging unphysically.  These trends are independent of the
final mass chosen, or the definition of mass used.  The EPS
formula as progenitor mass goes to zero also diverges, which we could not
approach due to our finite mass resolution, but the mass weighted
EPS calculation is finite at both ends\footnote{We thank Jun Zhang
for emphasizing this.}.
There is another notable difference between EPS and our simulation.
Though it is relatively small, our time step is still too large for all
mergers to be truly 2-body (see Fig.~\ref{fig:treeplot}), as implicitly
assumed by EPS.
A large fraction (20-50 per cent, depending on $M_f$) of the halos are
actually produced in 3 (or more)-body mergers.

Finally we also looked for evidence that recently merged halos clustered
differently than randomly chosen halos of the same mass.
The correlation function of 1:2 or 1:3 mergers appeared to be slightly
($<10\%$) enhanced at $1\,$Mpc compared to the random sample, but the
number of merged halos was too small for this to be statistically meaningful.
The effect thus appears to be modest, if present at all, just as was found
for lower redshift, high-mass halos \citep[e.g.][]{Got02,Per03,ScaTha03,Wet07}.
This suggests that the clustering of massive halos does not depend strongly
upon their recent merger history.  This in turn significantly eases the
modeling of merger-related processes, such as enhanced photon production
during reionization which we now discuss.

\section{Reionization effects} \label{sec:reion}

The rate of photon production in a galaxy can be enhanced by mergers,
which can trigger starbursts or possibly accretion onto a black hole
which may be present
\citep[e.g.][]{Car90,BarHer91,BarHer96,MihHer94,MihHer96,KauHae00,CavVit00}.
It is reasonable to anticipate that the mergers of large dark matter halos
could have similar effects on the photon production rate of the sources
within them.
We will make this assumption, and then consider the consequences of 
the merger rates computed above for the photon production distribution.

We frame our discussion in terms of a simple but promising model for
reionization proposed by \citet{FurZalHer04a}, though our result is true
more generally.  
In these models, a halo of a given mass $m$ (in units of some reference mass)
is considered a source of photons with rate
\begin{equation}
  \frac{dn_\gamma}{dt}=\zeta_t(m)m \quad .
\label{eqn:dndt}
\end{equation}
Usually $\zeta_t$ is taken to be mass independent, scale as $m^{2/3}$ or
transition from $m^{2/3}$ to $m^0$ at $M\sim 10^{10}\,h^{-1}M_\odot$
(\citealt{FurMcQHer06}, motivated by \citealt{Kau03}). 
A region around these halos is taken to be ionized if the photons within it
are sufficient to ionize all the interior mass.  Some extensions also
give recombinations spatial and/or temporal dependence
and incorporate this into finding the bubble properties 
\citep{FurOh05,FurMcQHer06,CohCha07}, or incorporate Eq.~(\ref{eqn:dndt})
into N-body simulations \citep{Ili06a,McQ07,Zah07}.
Under these assumptions the morphology of ionized regions can be computed from
the photon production rate and spatial distribution of dark matter halos.

A first step at including halo mergers within the above formalism (and its
generalizations) was presented in \citet{CohCha07}.
Those calculations were based on the Press-Schechter formalism, and so could
only provide average numbers of mergers for halos in a given mass range;
scatter was computed by assuming that the mergers had a Poisson distribution.
With our simulations we are able to check these assumptions and significantly
extend this work because we have access to the detailed merger history of each
halo.  This allows us to go beyond their analytic estimates to
explicitly calculate the full distribution of photon
production for a halo of mass $m$, taking into account the distribution of
histories and their associated (and different)
photon production rates for a fixed $m$.

{}From the merger tree for each halo at $z=10$ ($t=t_{\rm obs}$) we identify
which progenitors had at least one major merger (greater than 1:3 or 1:10),
and the time $t_{\rm merge}$ they occurred.
We include all of the mergers in the tree and we place $t_{\rm merge}$ at random
within the $10\,$Myr interval between the relevant outputs.
Each of these mergers is allowed to contribute ``excess'' photons beyond those
which would automatically be assigned to the halo on the basis of its $z=10$
mass, $M_h$, but the number of photons contributed is exponentially attenuated
with an $e$-folding time $\tau$.
The ``excess'' photon production is thus proportional to
\begin{equation}
  {\mathcal M}_s^\alpha \equiv \sum_{\rm merge} M^\alpha
     e^{(t_{\rm merge}-t_{\rm obs})/\tau} \quad ,
\end{equation}
where the sum is over all halos which have undergone a major merger and we
take $\alpha=1$ or $5/3$.  The exponential decay is motivated by modeling of
starbursts, e.g.~\citet{Con06}, hence the subscript $s$.
We also consider another variant, including all halos with major mergers
within $\tau$ of $t_{\rm obs}$, with no attenuation:
\begin{equation}
  {\mathcal M}_{bh} \equiv 
  \sum_{\rm merge} M \Theta(t_{\rm obs}-t_{\rm merge} -\tau) \quad ,
\end{equation}
where $\Theta(x) = 1$ if $x>0$, $1/2$ if $x=0$ and zero otherwise.
We denote this by a subscript $bh$, to indicate photon production by black
holes, which might have their photon production rate increase over time and
then decay once the fuel is exhausted.  Assuming a step-like function is a
crude first approximation to this uncertain physics.
In all cases we take the quiescent photon production to depend on the $z=10$
halo mass with the same index, $\alpha$, as $\mathcal{M}_s$.
We note this prescription
might cause some overcounting if many mergers occur within a short time
period and the gas becomes depleted from the earliest ones.
A more refined model would account for the evolving baryon budget within the 
halo, but our treatment is sufficient for the purpose of illustration.

The relative amplitudes of these two modes of photon production depend on
a number of different factors
(see e.g. \citet{CohCha07} for discussion and summary of estimates at these
redshifts)
but a factor $\beta\sim 5$ is not unreasonable for starbursts and
could be even larger for black holes.  The total photon production
is thus enhanced by a factor
\begin{equation}
  \varepsilon_{\rm mrg} \equiv
  \frac{M_h^\alpha + \beta{\mathcal M}_s^\alpha}{M_h^\alpha}
\label{eqn:enhance}
\end{equation}
for the ``starburst'' prescription, or its analogue
${\mathcal M}_s^\alpha \rightarrow {\mathcal M}_{bh}$ for the ``black hole''
prescription.  In principle both can contribute. 
We considered the two effects separately,
their combination is straightforward.  

\begin{figure}
\begin{center}
\resizebox{3.2in}{!}{\includegraphics{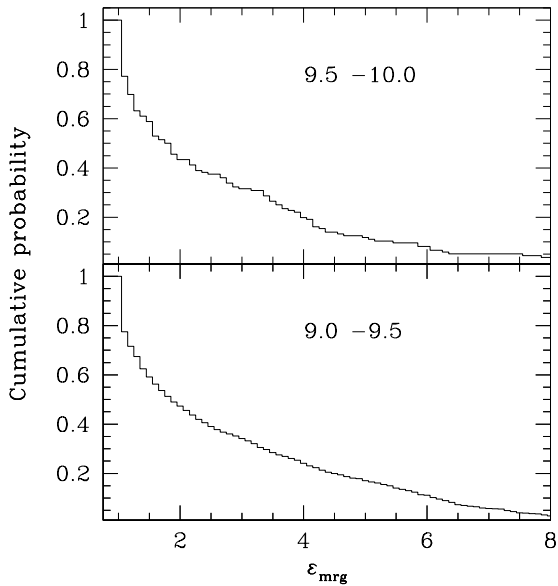}}
\end{center}
\caption{The cumulative probability for enhancement of photon production over
the quiescent case, $\varepsilon_{\rm mrg}$, for one toy model.  In this
example, mergers (1:3) enhance the quiescent photon production rate
($\sim m^{5/3}$) at $z=10$ with $\beta=5$ and $\tau=75\,$Myr (see text).
Roughly 80 per cent of these halos have some enhancement, shown for 2 bins in
halo mass (with the range in $\log_{10}(M/h^{-1}M_\odot)$ shown in each panel).
There are 1195 and 136 halos in the low- and high-mass bins, respectively.
The peak at $\varepsilon_{\rm mrg}=1$ comes from the 20 per cent of the halos
which have had no 1:3 mergers in the previous $140\,$Myr.
Trends for other parameters and parameterizations are discussed in the text.}
\label{fig:enhance}
\end{figure}

Figure \ref{fig:enhance} shows a typical example of the cumulative distribution
of enhancement factors, Eq.~(\ref{eqn:enhance}).  We took the starburst form,
1:3 mergers, $\alpha=5/3$, $\tau=75\,$Myr and $\beta=5$, but other cases are
very similar.  
The enhancement distribution is extended, with a long tail to high
$\varepsilon_{\rm mrg}$ and a peak at those halos which have not merged.
About half of the halos have twice the photon production, while 20 per cent
have no enhancement.
Choosing a larger $\beta$ increases the size of the enhancement, but does
not qualitatively change the form of the distribution.
Similarly, changing $\alpha$ or $\tau$ changes the detailed form of the
distribution but not its character.
Halos down to $10^8\,h^{-1}M_\odot$ show a very similar distribution of
enhancements.  By contrast the model for black hole accretion produces a
bimodal distribution, as the ``early'' mergers contribute relatively more
than in the case of the starbursts, leading to a second peak.

Even though the scatter in photon contributions from halo to halo is large
for a given mass, if a large number of such halos are found in a bubble,
their contributions to the photon numbers will tend to the mean, allowing
the distribution to be replaced by the average.
Precisely counting the number of halos of a given mass and the combined photon
scatter inside a typical bubble is unfortunately self-referential: changing
the ionization properties (including mergers) changes the bubble sizes and
thus the number of halos within.
Different assumptions about the nature of the sources and their feedback
can give drastically different bubble sizes, and the relative importance
of high vs.~lower mass halos \citep[e.g.][]{McQ07,Zah07}.
Given these uncertainties we consider properties in an average volume,
for illustration.

For quiescent photon production and $\zeta\propto m^{2/3}$, analytic estimates
such as Press-Schechter give that halos with $M>10^9\,h^{-1}M_\odot$ contribute
between $\frac{1}{3}-\frac{2}{3}$ of all photons.
Even choosing $\zeta\propto m^0$, such halos contribute $\sim$10 per cent of the
photons.  The number density of such halos is $\sim 0.03\,h^{3}{\rm Mpc}^{-3}$.
Bubble radii in different models range over several orders of magnitude.
A middle-of-the-road estimate is $3\,h^{-1}$Mpc, which would contain about 3
halos with $M>10^9\,h^{-1}M_\odot$.
The bubble radius would also be larger than the correlation length of our
halos, so clustering is only expected to change this number by a factor of
order unity.
A small number of halos contributing a large fraction of the photons means
that scatter in their photon production should affect the properties of the
bubbles.

Our calculation is relatively crude, but it suggests that the inclusion
of mergers into a more refined model of reionization could alter the
distribution of ionized regions.
For models based on approximate dynamics \citep[e.g.][]{McQ07,MesFur07,Zah07},
a possible first step would be to assign a merger history to the sources
at random.  This is accurate to the extent that recently merged halos are
not spatially biased with respect to a random sample of halos of the same
mass.  For models which marry the analytic model to dark matter
simulations\footnote{Unfortunately our simulation volumes are too small to
provide converged answers for this step with the existing runs.} the merger
history is known, so only the photon production rate needs to be modified.
More complex simulations involving radiative transfer will need
to follow the photon production history as the halos evolve, perhaps using
a semi-analytic model \citep[such as in][]{BNSL01,CiaStoWhi03,BNSL06}.
A full-blown simulation including radiative transfer and N-body in a large
enough volume is still out of reach
\citep[but see][for recent progress]{Sok03,KohGneHam05,TraCen06,Ili06a,Ili06b}.

\section{Summary and Conclusions} \label{sec:conclusions}

Using 5 N-body simulations with different sized boxes and particle loads
we considered the abundance, clustering and assembly histories of high
mass halos at high redshift.  We present results specifically for $z=10$,
but the evolution of the populations is smooth and the results will be
similar at slightly higher and lower redshift.
Like the halos of rich groups or clusters today the halos we consider are
in the process of forming, growing rapidly through accretions and mergers.
We found that they had larger velocity dispersions than a naive application
of the virial theorem would predict, due to a surface pressure from infalling
material.  Being recently formed, the halos were not very centrally
concentrated, leading to a factor of two difference between FoF(0.2) masses
and $M_{180}$.   
When measured against $M_{180}$ we found our halo abundances were closer to
the Press-Schechter fitting formula than that of \citet{SheTor99}, though the
simulations had more high mass halos than the analytic form.
If FoF(0.2) masses were used instead, the mass function approached that of
\citet{SheTor99}, in agreement with earlier work.
This discrepancy indicates that analytic models which assign an observable
to halos of a given size need to pay particular attention to the marker of
halo size employed.

The high mass halos were significantly clustered, and we calculated the
halo bias by taking the halo-mass cross correlation and dividing by the
matter auto-correlation function.  Our rare halos were more clustered than
the recent models of \citet{SheTor99,SheMoTor01,Tin05} and closer to the
models of \citet{Efs88,ColKai89,MoWhi96,Jin98}.

Merging is common, though not ubiquitous, in high mass halos at $z=10$.
Major mergers, with progenitor mass ratios less than 1:3, occurred
within $140\,$Myr of $z=10$ for more than half of halos with
$M>10^9\,h^{-1}M_\odot$.  We looked at the fraction of halos undergoing
mergers for a variety of lookback times and progenitor ratios, finding
more mergers for more massive halos, longer lookback times or less extreme
merger events.  Mass gains, parameterized by $m_f/m_i$, showed similar
trends even though not all merger events were two body within our $10\,$Myr
time step.  The EPS model provides a reasonable description of the
progenitor mass distribution, though it underpredicts the number of low
mass progenitors and diverges as $M_{\rm prog}\to M_h$.
The mass accretions histories predicted by EPS, as calculated by
\citet{Mil06}, provide only a qualitative guide to the mean mass accretion
histories seen in our simulations.

At $z=10$ reionization is expected to be underway due to photon production
{}from astrophysical objects which formed in collapsed halos.  Within the
context of a simple model which associates mergers with an increase in photon
production rate the photon production distribution developed a high rate tail
due to recently merged halos.
Including these photon production enhancements will likely drive scatter in
photon production at fixed halo mass.  Since the number of massive halos
within a typical ionized bubble can be small, this scatter in photon production
could well translate into additional scatter in bubble sizes and it would be
very interesting to include this effect in approximate models of reionization.
If the recently merged halos are not spatially biased with respect to other
halos of the same mass, including these effects in models, even those without
merger trees, should be straightforward.

JDC would like to thank D.~Holz and S.~Koushiappas for conversations.
We thank O.~Zahn for helpful comments on an early draft, and 
S.~Furlanetto, C.P.~Ma,
L.~Miller, E.~Neistein and J.~Zhang for comments on the final draft.
The simulations were analyzed on the supercomputers at the National Energy
Research Scientific Computing center.  MW was supported by NASA.
We thank the referee for encouraging us to run the $50\,h^{-1}$Mpc box.

\section*{Appendix}

We compared the progenitor mass distribution from the simulations with
the predictions of the extended Press-Schechter formalism.  To compute
the distribution in this formalism we need the number of progenitors
with $m_{i1}<m_i<m_{i2}$ at time $t-\Delta t$ for a given range of final
masses, $m_{f1}<m_f<m_{f2}$, at time $t$.  The EPS theory gives
\begin{eqnarray}
  \frac{N(m_{\rm prog},\Delta t)}{\Delta t} &=&
  \int_{m_{f1}}^{m_{f2}} dM\ N(M,t) \int_{m_{i1}}^{m_{i2}} dM_i
  \nonumber \\
  &\times&(M/M_i)\,\dot{P}_1(M_i\rightarrow M,t)
\label{eqn:EPS}
\end{eqnarray}
where the number of halos of mass $M$, $N(M,t)$, is given by the usual
mass function \citep{PreSch74}:
\begin{equation}
N(M,t) = \frac{1}{\sqrt{2\pi}} \frac{\rho_0}{M}
  \frac{\delta_c(t)}{\sigma^3(M)}
  \left|\frac{d\sigma^2}{dM}\right|
  \exp\left[-\frac{\delta_c^2(t)}{2\sigma^2(M)}\right]
\end{equation}
with $\rho_0$ the background density.  To get Eq.~(\ref{eqn:EPS}) the number
of halos at a given time is multiplied by the fraction of halos that have mass
$M$ at $t$ and have jumped from mass $M_i$ within $\delta t$:
\begin{equation}
  \begin{array}{l}
  (M/M_i)\dot{P}_1(M_i\to M;t) dM_i\delta t = \\
  \; \; \; (M/M_i)(2 \pi)^{-1/2} (\Delta\sigma^2)^{-3/2}
  [-\frac{d \delta_c(t)}{dt} ] |\frac{d\sigma^2(M_i)}{d M_i}|
   dM_i \delta t \; .
\end{array}
\label{eq:pdot}
\end{equation}
where $\Delta\sigma^2=\sigma(M_i)^2-\sigma(M)^2$.
To produce the EPS predictions discussed in \S\ref{sec:merge}, we take
$\delta t=\Delta t=10\,$Myr, assuming that this time is small.

\end{document}